\documentstyle[12pt,a4,psfig]{article}
\begin{document}

\newcommand{\E}[1]{$\cdot 10^{#1}$}

\renewcommand{\baselinestretch}{1.5}

\title{The structure of the atomic helium trimers:\\ Halos and Efimov states}

\author{E. Nielsen, D.V. Fedorov and A.S.~Jensen  \\
Institute of Physics and Astronomy, \\
Aarhus University, DK-8000 Aarhus C, Denmark }

\maketitle

\begin{abstract}
The Faddeev equations for the atomic helium-trimer systems are solved
numerically with high accuracy both for the most sophisticated
realistic potentials available and for simple phenomenological
potentials. An efficient numerical procedure is described. The
large-distance asymptotic behavior, crucial for weakly bound
three-body systems, is described almost analytically for arbitrary
potentials. The Efimov effect is especially considered. The geometric
structures of the bound states are quantitatively investigated. The
accuracy of the schematic models and previous computations is
comparable, i.e. within 20\% for the spatially extended states and
within 40\% for the smaller $^4$He-trimer ground state. \\

\end{abstract}

\newpage

\section{Introduction}

Weakly bound two- and three-body systems in low angular momentum
states are spatially very extended \cite{rii92,fed94a,goy95,ric94}. We
shall call them halos, if a substantial fraction of the wave function
is outside the range of the potentials. Each of the constituent
``particles'' may be a composite system, provided it is a tightly
bound structure with a binding energy large compared to the
interparticle binding energy. Then the corresponding (intrinsic and
halo) degrees of freedom decouple and a few-body treatment is
appropriate. Prominant examples are the new class of nuclear states
called nuclear halos \cite{han95,tan96} and various molecular systems,
in particular the atomic helium trimer \cite{esr96,efi70}.

Halo states are in general characterized by their large spatial
extension. They could as well appear as excited states of systems
which in the ground state are of a different nature and spatially
confined. Borromean systems, where no binary subsystem is bound, are
particularly interesting three-body halo candidates \cite{zhu93}. They
have by definition a relatively low binding energy. Many nuclear
examples are available \cite{fed94a,han95,tan96} and until recently it
was unclear whether the helium trimer was of Borromean nature
\cite{esr96}.  The halo concept is general and of interest in
different subfields of physics
\cite{rii92,fed94a,goy95,ric94,han95,tan96,rii94,esr96,efi70,zhu93,
ric92,lin95}. In
the simplest case of two-body halo systems universal relations between
size and binding energy were discussed \cite{rii92,mis97}. Also the
general properties of three-body halos are intensively discussed
\cite{fed94a,goy95,ric94,han95,fed93b}

The halo structure is almost entirely determined by the tail
properties of the effective interactions between the particles. If the
finite-size particles have a substantially overlap, the halo degrees
of freedom do not decouple and the few-body approximations are
inaccurate \cite{fed94a,fed93b}.  The effective two-body interactions
should then reproduce low-energy scattering properties, i.e. phase
shifts or it may be sufficient with the correct scattering length and
effective range \cite{fed94a,fed93b,cob97a}. For molecular systems it
is possible to compute the two-body effective potentials directly from
the electromagnetic interaction \cite{esr96,azi91}.

The halo structure is technically difficult to compute accurately,
since precise knowledge of the behavior of the effective radial
(three-body) potential at large-distance is indispensable and at the
same time the small distances, responsible for the actual size of the
binding energy, must also be properly included.  Fortunately, a method
treating the large distances analytically and the short distances
numerically has recently become available
\cite{cob97a,fed93,cob97b,jen97}.  The method is very powerful as
demonstrated by the succesful investigation of the patological Efimov
effect \cite{efi70,fed93,fed94c}. Within the method the Faddeev
equations are solved in coordinate space in two steps. First the
angular part is solved for each hyperradius, thereby providing the
adiabatic basis. This is usually the most difficult part, where
semianalytic large-distance asymptotic solutions are
employed. Afterwards the coupled set of radial equations for the
expansion coefficients is solved.

Another somewhat similar method was recently used to study the helium
trimer \cite{esr96}. This method exploit the fact that the interaction
(in addition to two distance-coordinates) only depends on one angle,
since the three Euler angles describe rotations of the system as a
whole. In principle the same adiabatic basis is then
obtained. However, the computations only include the lowest angular
basis state which already seems to involve extensive calculations.

The $^4$He-dimer and trimer systems have recently been experimentally
established as weakly bound systems although without direct
measurement of the binding energies \cite{sch96}. However, the size of
the $^4$He-dimer was found to be exceedingly large compared to the
range of the two-body potential. In the conceptually simple experiment
transmission was measured through a screen consisting of holes with
radii comparable to the dimer size \cite{lou96}.  The binding energy
of the dimer must consequently be exceedingly small. However, this
does not constrain the binding energy of the three-body system which,
due to the Thomas-effect \cite{tho35}, still can be much larger than
the binding energy of the two-body system. On the other hand, results
from solving the Schr\"{o}dinger equation as well as momentum space
Faddeev calculations strongly suggest that the trimer binding energy
indeed is very small \cite{esr96,glo86}.  Both computations conclude
that the trimer has two bound states of which the excited state has
the characteristica of an Efimov state: When the attraction is
increased it disappears into the continuum away from the discrete
three-body spectrum.

Interesting systems can be formed by substituting one or more of the
$^4$He-atoms by $^3$He-atoms. The electron structure is identical and
the effective two-body interaction is then unchanged. Due to the
smaller masses the three-body states move up in energy and
calculations predict that ${^4}$He$^3$He$_2$ and the
$^3$He-trimer both are unbound \cite{esr96}. There is no simple
relation between the size and the binding energy of these systems and
detailed calculations of the three-body wave functions are needed.

The purpose of the present paper is three-fold. First, we shall repete
the investigation of Esry et al. \cite{esr96} and extent it to include
sufficiently many adiabatic basis states to obtain precise
solutions. We shall keep an eye on the possible occurrence of Efimov
states.  Second, we shall define simple two-body potentials
reproducing the low-energy scattering behavior of the realistic
potentials. We shall assess the accuracy of such approximation
schemes, which reduce the necessary computational efforts by orders of
magnitude. Third, we shall describe an improved method able to deal
efficiently with large repulsive cores as traditionally used in the
parametrization of the two-body molecular interaction.

After the introduction we describe first briefly in section 2 the
general method of solving the Faddeev equations using hyperspherical
coordinates.  The angular equations and the effective radial potential
at large distances are then analyzed in sections 3 and 4. The rather
delicate numerical procedure and the corresponding results are
discussed in sections 5 and 6, respectively. Finally section 7
contains the summary and the conclusions.

\section{Theoretical method}

We shall consider an angular momentum zero system of three interacting
inert ``particles'', where the $k$'th particle has mass $m_{k}$ and
coordinate ${\bf r}_k$. Their intrinsic degrees of freedom are frozen
and only the three-body degrees of freedom shall be treated here.  The
two-body interactions between the particles $i$ and $j$ are $V_{ij}$.
We shall use the three sets of Jacobi coordinates ${\bf x}_i$ and
${\bf y}_i$ and the corresponding three sets of hyperspherical
coordinates $(\rho,\Omega_{i})$=($\rho$, $\alpha_i$, $\Omega_{x_i}$,
$\Omega_{y_i}$) defined by \cite{fed94a,zhu93}
\begin{eqnarray}\label{ea1}
 {\bf x}_i = {\mu_{jk}} {\bf r}_{jk} \; , & & \;
 {\bf r}_{jk} = {\bf r}_j-{\bf r}_k \; ,   \\
   {\bf y}_i = {\mu_{i(jk)}} {\bf r}_{i(jk)} \; ,  & &  \;
 {\bf r}_{i(jk)} = {\bf r}_i - \frac{m_j {\bf r}_j+m_k {\bf r}_k}{m_j+m_k} \;,
 \\
 x_i = \rho \sin \alpha_i \; , & & \; y_i = \rho \cos \alpha_i \;,  \\
 \mu_{jk} = \left( \frac{1}{m}\frac{m_j m_k}{m_j+m_k} \right)^{1/2} \; , & & \;
 \mu_{i(jk)} = \left( \frac{1}{m}\frac{m_i(m_j+m_k)}{m_1+m_2+m_3} 
 \right)^{1/2} \; ,
\end{eqnarray}
where $\{i,j,k\}$ is a cyclic permutation of $\{1,2,3\}$ and $\mu^2$
are the reduced masses of the subsystems in units of a normalization
mass $m$.  The volume element in terms of one of the sets of hyperspherical
coordinates is given by $\rho^5{\rm d}\Omega{\rm d}\rho$, where ${\rm
d}\Omega=\sin^2 \alpha \cos^2 \alpha {\rm d}\alpha {\rm d}\Omega_x
{\rm d}\Omega_y$. The kinetic energy operator is
\begin{eqnarray} \label{e10}
T=\frac{\hbar ^{2}}{2m}\left( -\rho ^{-5/2}\frac{\partial ^{2}}{\partial
\rho ^{2}}\rho ^{5/2}+\frac{15}{4\rho ^{2}}+\frac{\hat{\Lambda}^{2}}{\rho
^{2}}\right) \; , \\ \label{e15}
	\hat \Lambda^2=-{1 \over \sin(2\alpha)}
      {{\partial}^2\over {\partial} \alpha^2} \sin(2\alpha)  
	+{\hat l_x^2 \over {\sin^2 \alpha}}
 	+{\hat l_y^2 \over {\cos^2 \alpha}} -4  \; , 
\end{eqnarray}
where the angular momentum operators $\hat l_{x}^2$ and $\hat l_{y}^2$
are related to the ${\bf x}$ and ${\bf y}$ degrees of freedom.

The total wave function $\Psi$ of the three-body system is written as
a sum of three components $\psi^{(i)}$, which in turn for each
$\rho$ are expanded on a complete set of generalized angular functions
$\Phi^{(i)}_{n}(\rho,\Omega_i)$
\begin{equation} \label{e20}
\Psi = \sum_{i=1}^3 \psi^{(i)}(\mbox{\bf x}_i, \mbox{\bf y}_i)
= \frac {1}{\rho^{5/2}} 
  \sum_n f_n(\rho)  
\sum_{i=1}^3 \Phi^{(i)}_{n}(\rho ,\Omega_i)
= \frac {1}{\rho^{5/2}} 
  \sum_n f_n(\rho)  \Phi_{n}(\rho ,\Omega) \; ,
\end{equation}
where $\rho^{-5/2}$ is the radial phase space factor and $n=1,2,3...$

The angular functions, the adiabatic basis, are now for each $\rho$
chosen as the eigenfunctions of the angular part of the Faddeev
equations:
\begin{equation} \label{e25}
{\hbar^2 \over 2m}\frac{1}{\rho^2} \left(\hat\Lambda^2-\lambda_n(\rho)\right)
\Phi^{(i)}_{n} 
 +V_{jk} (\Phi^{(i)}_{n}+\Phi^{(j)}_{n}+
                \Phi^{(k)}_{n}) = 0   \; ,
\end{equation}
where $\{i,j,k\}$ is a cyclic permutation of $\{1,2,3\}$.

The radial expansion coefficients $f_n(\rho)$ are obtained from 
a coupled set of ``radial'' differential equations \cite{fed94a}, i.e.
\begin{eqnarray} \label{e30}
   \left(-\frac{\partial ^{2}}{\partial\rho ^{2}}
   -{2m(E-V_3(\rho))\over\hbar^2}+ \frac{1}{\rho^2}\left( \lambda_n(\rho) +
  \frac{15}{4}\right)  -Q_{n n} \right)f_n(\rho)
 \nonumber  \\
  = \sum_{n'\neq n}   \left( 2P_{n n'}\frac{\partial}{\partial\rho }
   + Q_{n n'} \right)f_{n'}(\rho) \; ,
\end{eqnarray}
where $E$ is the three-body energy, $V_3(\rho)$ is an anticipated
additional three-body potential and the functions $P$ and $Q$ are
defined as angular integrals:
\begin{equation}\label{e40}
   P_{n n'}(\rho)\equiv
   \int d\Omega \Phi_n^{\ast}(\rho,\Omega)
   {\partial\over\partial\rho}\Phi_{n'}(\rho,\Omega)  \; ,
\end{equation}
\begin{equation} \label{e45}
   Q_{n n'}(\rho)\equiv 
   \int d\Omega \Phi_n^{\ast}(\rho,\Omega)
   {\partial^2\over\partial\rho^2}\Phi_{n'}(\rho,\Omega)  \; .
\end{equation}

Explicit expressions for these coupling terms can be obtained by
deriving eq.(\ref{e25}) with respect to $\rho$. By use of the
normalization condition $\left<\Phi_n | \Phi_{n'}\right>\equiv\delta_{nn'}$.
we obtain
\begin{eqnarray} \label{e49}
& & P_{nn}=0 \; , \;
 P_{nn'} = -{\left<\Phi_{n}\left|{\partial (\rho^2 v)\over \partial \rho}
\right| \Phi_{n'}\right>\over
	\lambda_n-\lambda_{n'}} \;\; {\rm for} \;\; \hbox{$n\not=n'$} \; , \\
& &  Q_{nn}=\sum_{m\not=n}P_{nm}P_{mn} \; ,\\
& &  Q_{nn'}= \sum_{m\not=n,m\not=n'}{\lambda_n-\lambda_m\over
\lambda_n-\lambda_{n'}}P_{nm}P_{mn'}  \\ \nonumber  & & -
{\left<\Phi_{n}\left|{\partial^2(\rho^2 v)\over{\partial}^2\rho}
\right|\Phi_{n'}\right>
   +2\left(\left<\Phi_{n}\left|{\partial (\rho^2 v)\over \partial\rho}\right|
\Phi_n\right>-
   \left<\Phi_{n'}\left|{\partial (\rho^2 v)\over\partial\rho}\right|
\Phi_{n'}\right>\right)P_{nn'}
\over
	\lambda_n-\lambda_{n'}} \;\; {\rm for} \;\; \hbox{$n\not=n'$} \; ,
\end{eqnarray}
where $v$ essentially is the sum of the three two-body potentials,
i.e. $v =  \sum_{i=1}^{3} v_i$ with $v_i(x) = 2m
V_{jk}(x/\mu_{jk}) / \hbar^2$. Thus $P$ is antisymmetric.

\section{Large-distance angular solution}

The angular functions $\Phi^{(i)}_{n}(\rho ,\Omega_i)$ are expanded in
products of the three-body spherical harmonics $Y_{\ell_x
m_x}(\Omega_{x_i})$ and $Y_{\ell_y m_y}(\Omega_{y_i})$ where the
orbital angular momenta and their projections associated with {\bf x}
and {\bf y} are $(\ell_x, m_x)$ and $(\ell_y, m_y)$. The total
angular momentum is zero and we get
\begin{eqnarray} \label{e60}
\Phi^{(i)}_{n}(\rho ,\Omega_i) = \sum_{\ell}
\frac{\phi_{n \ell}^{(i)}(\rho, \alpha_i)}{\sin(2\alpha_i)}
 \left[Y_{\ell m}(\Omega_{x_i}) \otimes Y_{\ell -m}(\Omega_{y_i})
\right]^{00}  \; ,
\end{eqnarray} 
where $\sin(2\alpha_i)$ is a factor related to phase space. We can now 
insertion of eq.(\ref{e60}) into eq.(\ref{e25}) and obtain  
an equation for each of $\ell$ projections of each Faddeev component:
\begin{eqnarray} \label{e85}
  \left(-\frac{\partial^2 } {\partial \alpha_i^2} + 
\frac{\ell(\ell+1)}{\sin^2 \alpha_i} 
+ \frac{\ell(\ell+1)}{\cos^2 \alpha_i} 
+ \rho^2 v_i(\rho \sin{\alpha_i})  - \nu^2_n(\rho)\right)
  \phi_{n \ell}^{(i) }(\rho, \alpha_i) = \\ \nonumber
-\rho^2 v_i(\rho \sin{\alpha_i}) R^{(i)}_{\ell}(\alpha_i) \; ,
\end{eqnarray}
where 
$R^{(i)}_{\ell}$ is the sum of the other two 
Faddeev components projected onto
$\left[Y_{\ell m}(\Omega_{x_i}) \otimes Y_{\ell -m}(\Omega_{y_i})
\right]^{00}$ in the $i$'th Jacobyan system. I.e. $R^{(i)}_{\ell}$ is a
sum of all the non-zero $\ell$ -contributions of the two other 
Faddeev components. A detailed analysis shows that
$R^{(i)}_{\ell}(\alpha_i)$ is proportional to $\sin^{\ell+1}\alpha_i$.

When $\rho$ is large the
potential is zero everywhere except for small $\alpha_i$. Thus for $\ell>0$
the effect of the potential on the left hand side of the equation 
have to been seen through a centrifugal barrier proportional
to $\ell(\ell+1)$ and the effect of the
potential on the right hand side is quenched by a factor of 
$\alpha_i^{1+\ell}$.

Thus for non-zero $\ell$ all terms containing the potential can be put to
zero for large enough $\rho$. Thus the free equation appear. So if any
of the Faddeev components have a finite contribution with $\ell>0$ for 
$\rho\rightarrow\infty$ we obtain
the free solution for that component 
with the free eigenvalue $\nu_n^2=(K+2)^2$ where
$K$ is an even integer equal or greater than $2\ell$. Again a detailed
analyses shows that $\nu_n^2=(K+2)^2+O(\rho^{-1-2\ell})$. This analysis
is actually quite similar to the ones given below except higher 
transcendental functions have to be used in the derivation.
 
If we now assume that all the Faddeev components only have contributions from
$\ell=0$ for $\rho\rightarrow\infty$ we end up with three coupled 
asymptotic equations
\begin{eqnarray} \label{e87}
 \left(\frac{\partial^2 } {\partial \alpha_i^2} 
 - \rho^2 v_i(\rho \sin{\alpha_i}) + \nu^2(\rho) \right)
  \phi_{0 }^{(i) }(\rho, \alpha_i)  =  \rho^2 v_i(\rho \sin{\alpha_i})
 \nonumber \\
 \times \sum_{j \neq i} \frac{1}{\sin(2\varphi_k)} 
\int_{|\varphi_k-\alpha_i|}^{\pi/2-|\pi/2-\varphi_k-\alpha_i|} d\alpha_j
   \phi_{ 0 }^{(j) }(\rho, \alpha_j) \; ,
\end{eqnarray}
\begin{equation} \label{e97}
 \tan \varphi_{k} = \sqrt{\frac{m_k(m_i+m_j+m_k)}{m_i m_j}}  \; .
\end{equation}
The short-range potential $v_i(\rho \sin{\alpha_i})$ can be assumed to
vanish for $\alpha_i$ larger than a constant $\alpha^{(i)}_0$ which
for large distances must decrease inversely proportional to $\rho$.
The corresponding solution to eq.(\ref{e87}) is then
\begin{equation}\label{e53}
 \phi_{0}^{(i)}(\rho,\alpha_i) = A_{0}^{(i)}
\sin\left[ \nu \left(\alpha_i - \frac{\pi}{2} \right) \right] \; 
  {\rm for} \; \alpha_i > \alpha^{(i)}_0 \; , 
\end{equation}
where $A_{0}^{(i)}$ is an arbitrary constants and the proper boundary
condition $\phi_{0}^{(i)}(\rho,\alpha=\pi/2)=0$ explicitly is
included.  

The function $\phi_{0 }^{(j)}(\rho, \alpha_j)$ in eq.(\ref{e87}) is
only needed when $v_i(\rho \sin{\alpha_i})$ remains finite, i.e. for
$\alpha_i < \alpha^{(i)}_0$. When $\alpha^{(i)}_0 + \alpha^{(j)}_0 <
\varphi_k$ the integration in eq.(\ref{e87}) only includes the region
where eq.(\ref{e87}) is valid. Therefore as soon as $\rho$ corresponds
to distances outside the short-range potentials we can rewrite
eq.(\ref{e87}) as
\begin{eqnarray} \label{e93}
 \left(\frac{\partial^2 } {\partial \alpha_i^2} 
- \rho^2 v_i(\rho \sin{\alpha_i}) + \nu^2(\rho) \right)
  \phi_{0 }^{(i) }(\rho, \alpha_i)  = 2 \rho^2 
 v_i(\rho \sin{\alpha_i}) \frac{\sin(\nu\alpha_i)}{\nu}  \nonumber \\
 \times  \sum_{j \neq i}  A_{0}^{(j)} \frac{
 \sin\left(\nu(\varphi_{k}-{\pi\over 2})\right)}{\sin(2\varphi_k)}   \; .
\end{eqnarray}

The solution to eq.(\ref{e93}) is given in eq.(\ref{e53}) for 
$\alpha_i > \alpha^{(i)}_0$ and for $\alpha_i < \alpha^{(i)}_0$ we find
\begin{equation}\label{e94}
\phi_{0 }^{(i) }(\rho, \alpha_i)  = \phi_{0h}^{(i)}(\rho, \alpha_i)
 - \frac{2}{\nu} \sin(\nu \alpha_i) 
 \sum_{j \neq i}  \frac{A_{0}^{(j)} 
 \sin\left(\nu(\varphi_{k}-{\pi\over 2})\right)}{\sin(2\varphi_k)} \; , 
\end{equation}
where $\phi_{0h}^{(i)}$ is the solution to the homogeneous equation
\begin{equation}\label{e96}
 \left(\frac{\partial^2 } {\partial \alpha_i^2} 
 - \rho^2 v_i(\rho \sin{\alpha_i})  + \nu^2(\rho) \right) 
 \phi_{0h}^{(i)}(\rho, \alpha_i) = 0 \; 
\end{equation}
with the boundary conditions $\phi^{(i)}_{0h}(\rho,\alpha_i={\pi\over
2})= \phi^{(i)}_{0h}(\rho,\alpha_i=0) = 0$. If $v_i(\rho
\sin{\alpha_i})$ varies within the interval $\alpha_i <
\alpha^{(i)}_0$ the solution to eq.(\ref{e96}) could be complicated
and only obtained numerically. However, if $v_i(x)$ approaches a
finite constant for small $x$ we find immediately
\begin{equation}\label{e99}
\phi_{0h}^{(i)}(\rho, \alpha_i) = B^{(i)}_{0}
  \sin\left(\alpha_i \sqrt{ \nu^2(\rho) - \rho^2 v_i(0)} \right) 
\end{equation}
for arbitrary constants $B^{(i)}_{0}$.

Assuming that $\nu\alpha^{(i)}_0$ is small for large $\rho$ we can
expand the right hand side of eq.(\ref{e87}) to leading order in
$\alpha_i$.  This gives
\begin{equation} \label{e90}
 \left(\frac{\partial^2 } {\partial \alpha_i^2} 
- \rho^2 v_i(\rho \sin{\alpha_i}) + \nu^2(\rho) \right)
  \phi_{0 }^{(i) }(\rho, \alpha_i)
     =  2\alpha _i   C^{(i)}
  \rho^2 v_i(\rho \sin{\alpha_i}) \; ,
\end{equation}
\begin{equation} \label{e95}
 C^{(i)} \equiv 
 \frac{\phi_{0}^{(j) }(\rho, \varphi_{k})} 
{\sin(2 \varphi_{k})} + 
\frac{\phi_{0}^{(k) }(\rho, \varphi_{j})} 
{\sin(2 \varphi_{j})}  \; ,
\end{equation}
The eigenvalue solutions $\nu^2_n$ to eqs.(\ref{e87}), (\ref{e93}) and
(\ref{e90}) converge towards the hyperspherical spectrum as $\rho$
increases. Due to the coupling the asymptotic values are now
approached over a distance defined by the scattering lengths, which
might be very much larger than the ranges of the interactions. 

The potentials $\rho^2 v_i(\rho \sin{\alpha_i})$ vanish for large
$\rho$ for all $\alpha_i$ except in a narrow region around zero. The
conditions for the effective range approximation therefore become
better and better fulfilled as $\rho$ increases and any potential with
the same scattering length and effective range would lead to the same
results.  Let us then in the region of large $\rho$ use square well
potentials $V_{jk}(r) = - S^{(i)}_0 \Theta(r<R_i)$, or equivalently
expressed by the reduced quantities $v_{i}(x) = - s^{(i)}_0
\Theta(x<X_i=R_i \mu_{jk})$, where the parameters are adjusted to
reproduce the two-body scattering lengths and effective ranges of the
initial potential. The corresponding solutions are then accurate
approximations to our original problem at distances comparable to the
range of the potentials \cite{jen97}.

Using square well potentials we can now easily define the constants
$\alpha^{(i)}_0 = \arcsin(X_i/\rho)$ such that the potentials
$v_i(\rho \sin{\alpha_i})$ for large $\rho$ are zero in region II
where $\alpha_i>\alpha^{(i)}_0$ and finite and constant in region I
where $\alpha_i < \alpha^{(i)}_0 \ll 1$. The square well solutions are
then given in eqs.(\ref{e53}), (\ref{e94}) and (\ref{e99}).  We can
also solve the approximate equation in eq.(\ref{e90}) for both region
I and II. We obtain again the same solutions, where $\nu^2(\rho)$ is
neglected compared to $\rho^2 v_i(0)$ and $\sin(\nu\alpha_i)$ in the
last term of eq.(\ref{e94}) should be replaced by
$\nu\alpha_i$. Matching these solutions at the boundary
$\alpha_i=\alpha^{(i)}_0$ provides the quantization condition and
determines the eigenvalue $\nu$.

Also the decoupled equations in eq.(\ref{e85}) for non-zero
$\ell$-values can be solved for square well potentials. In region I
and II the solutions are proportional to the spherical Bessel
functions and the hyperspherical polynomials, respectively.

All these square well solutions coincide at large distances with the
solutions for any general potential with the same scattering length
and effective range.  However, when the two-body potentials are
strongly repulsive at small distances it is an advantage to use
eq.(\ref{e93}) with the correct potential instead of both the square
well solution and the additional approximation in eq.(\ref{e90}).  A
faster convergence (for smaller $\rho$) towards the large-distance
asymptotics is then obtained.

The method described here is useful when the large-distance behavior
is important. In extreme cases as for example for Efimov states and
halo systems the large distances are necessary and otherwise difficult
or impossible to compute sufficiently accurate.  For large distances,
where the approximations become increasingly better, it is in any case
much easier to obtain accurate solutions from the asymptotic equations
than from the original integro-differential equations in
eq.(\ref{e87}). Thus we have formulated a convenient numerical
procedure.

\section{The effective potential and the Efimov states}

The most important ingredient in the radial equation is the angular
eigenvalue $\lambda=\nu^2-4$. We can extract the essential properties
by analytical analysis of a schematic model.  For simplicity we
consider three identical bosons. We first find the square well
potential $-S_0\Theta(r<R_0)$ with the same scattering length $a_s$
and effective range $R_{eff}$ as the original two-body potential. The
lowest-lying adiabatic radial potentials are then asymptotically close
for the two potentials for hyperradii $\rho$ larger than $2R_0$. The
large-distance behavior is essentially determined by the
$s$-waves. The resulting eigenvalue equation, derived from
eqs.(\ref{e53}), (\ref{e94}) and (\ref{e99}), is a simple
transcendental equation \cite{jen97}
\begin{eqnarray} \label{e1}
& \frac{\kappa}{\nu}  \cos( \alpha_0 \kappa)
  \left[ \frac{8}{\sqrt{3}} \sin( \nu \; \pi/6 ) 
  \sin(\nu \alpha_0 ) - \nu
 \sin \left((\alpha_0 - \pi/2) \nu \right) \right]
   \nonumber \\
& = \sin(\alpha_0 \kappa)
 \left[ \frac{8}{\sqrt{3}} \sin( \nu \; \pi/6 ) 
  \cos( \alpha_0 \nu) - \nu
 \cos \left((\alpha_0 - \pi/2) \nu \right) \right] \; ,
\end{eqnarray}
where $\kappa = \sqrt{2mS_0\rho^2/\hbar^2 + \nu^2(\rho)}$ and
$\alpha_0 = \arcsin(R_0 / \rho \sqrt{2} )$.

The lowest-lying hyperradial potential in eq.(\ref{e30}) are to the
lowest orders in $1/\rho$ obtained by expansion of eq.(\ref{e1}). For
$R_0 \ll \rho \ll |\mu_{ij}a_s|$, where $a_s$ is the scattering
length, we find 
\begin{equation} \label{e22}
\lambda(\rho) \approx
 \nu^2_E - 4 - c_E \frac{R_0\mu_{ij}}{\rho} 
 (1 - \frac{\pi^2}{48\nu^2_E})  \; ,
\end{equation}
\begin{equation} \label{e24}
c_E = \frac{  \nu^3_E
 \sin( \frac{\pi}{2} \nu_E ) }
 {  \frac{4 \pi}{3\sqrt{3}} \cos( \frac{\pi}{6} \nu_E ) 
 - \cos( \frac{\pi}{2} \nu_E )
  +  \frac{\pi}{2} \nu_E \sin( \frac{\pi}{2} \nu_E ) }
 = 0.486214  \; ,
\end{equation}
where $\nu^2_e=-1.0125$ is the solution to the Efimov equation
obtained by expansion of eq.(\ref{e1}) in the limit where $a_s/\rho=
\pm \infty$, i.e.
\begin{equation} \label{e26}
8 \sin( \nu_E \pi/6 ) =
\nu_E\sqrt{3} \cos (\nu_E \pi/2) \; .  
\end{equation}
In the other limit where $|\mu_{ij}a_s| \ll \rho$ we find instead for
the lowest eigenvalue\footnote{We assume a negative
scattering length for a two-body system with one bound state}
\begin{equation} \label{e28}
\lambda(\rho) \approx  - \frac{48a_s\mu_{ij}}{\pi\rho} \; \; 
{\rm for} \;\; a_s > 0 \;\; , \;\;
\lambda(\rho) \approx  - \frac{\rho^2}{\mu_{ij}^2a_s^2} \; \; 
{\rm for} \;\; a_s < 0 \;.
\end{equation}

The effective radial potential is in this way obtained analytically
for the square well potential for various distance intervals when the
scattering length is large compared to the interaction range. However,
as indicated above these results are more general, since they can be
expressed in terms of scattering length and effective range, which are
universal quantities characterizing low-energy two-body scattering
properties of short-range potentials. For distances smaller than $R_0$
the details of the original two-body potentials are essential and the
square well results are not applicable. This part of the potential is
decisive for the actual (small) binding energy of the three-body
system.

Between a few times $R_0$ and a few times below $15\mu_{ij}|a_s|$
eq.(\ref{e22}) is valid. This region could be called the Efimov
region, since the Efimov states appear here for $\rho<|\mu_{ij}a_s|$
where the radial potential in eq.(\ref{e30}) behaves as
$\rho^{-2}$. The lowest angular and radial wave functions $\Phi$ and
$f_n$ are then obtained from eqs.(\ref{e60}), (\ref{e53}) and
(\ref{e30}) as
\begin{equation} \label{e78}
\Phi_n \propto \sum_{i=1}^{3} \frac{\sin(\nu_E(\alpha_i-\pi/2))}
{\sin(2\alpha_i)} \;\; , \;\;
f_n \propto \sqrt{\rho} K_{\nu_E}\left(\rho\sqrt{2m|E|/\hbar^2}\right) \; ,
\end{equation}
where $K_{\nu_E}$ is the modified Bessel function with an imaginary
index.  The radial function could easily be obtained from
eq.(\ref{e30}) in the limits where either the energy or the
centrifugal barrier term is dominating. The same result is found by
expansion of eq.(\ref{e78}) for small and large values of $\rho$
\begin{eqnarray}  \label{e77}
 f_n \propto \sqrt{\rho} \sin{\left(|\nu_E| \ln (\frac{\rho}
{\mu_{ij}R_{eff}})\right)} \;\; {\rm for} \;\;
 \rho^2  < \hbar^2/(2m|E|) \; . \; \\ f_n \propto \exp(-\rho
 \sqrt{2m|E|/\hbar^2}) \;\; {\rm for} \;\; \rho^2 > \hbar^2/(2m|E|) 
 \; ,
\end{eqnarray}
where the zero point for the first oscillation in $\rho$ is assumed to
be $\mu_{ij}R_{eff}$.  The exponential fall off at large distance
occurs for all bound states for distances larger than
$\hbar^2/(2m|E|)$, since the effective radial potential falls off
faster than $\rho^{-3}$.

The number of Efimov states is determined by the number of possible
oscillations of the sine function within the interval
$\rho<|15\mu_{ij}a_s|$. Each new oscillation corresponds to another
further excited state and the zero point for the $k$'th oscillation is
then given by
\begin{equation} \label{e79}
\rho_k \approx \mu_{ij}R_{eff} \exp\left(\frac{k\pi}{|\nu_E|}\right) \; .
\end{equation}
The number of Efimov states N is estimated from
eq.(\ref{e28}) by $\rho_N \approx |15\mu_{ij}a_s|$, i.e.
\begin{equation} \label{e81}
 N \approx \frac{|\nu_E|}{\pi}\ln\left(\frac{15|a_s|}{R_{eff}}\right) \; .
\end{equation}
The $k$'th state has a radial extension of about $\rho_k$,
i.e. exponentially increasing with $k$, and correspondingly an
exponentially decreasing energy
\begin{equation} \label{e83}
 E_k \approx  \frac{\hbar^2}{2mR_{eff}^2} 
\exp\left(-\frac{2k\pi}{|\nu_E|}\right) \; .
\end{equation}
Clearly infinitely large $a_s$ then gives infinitely many bound Efimov
states. They are extremly weakly bound and extremely spatially
extended.

Efimov states may also be present in non-identical bosonic three-body
systems. The condition is that at least two of the binary subsystems
have bound $s$-states at zero energy or equivalently infinitely large
scattering lengths. The third binary subsystem must be without bound
states.  A finite number of states with similar properties appear when
the conditions are approximately fulfilled.  The appearance is
determined by the large-distance behavior of the zero angular momentum
states.

\section{Numerical procedure}

It should be straightforward to solve the equations of motion
numerically. However, two major difficulties must be tackled. The
first is related to the large short-range repulsive core where the
wave function must be extremely small.  The potential has an
attractive pocket at short distances where the wave function also is
small. These inner regions are where the potential is relatively
large, are decisive for the energy of the bound states and the
corresponding wave function must therefore be calculated with high
accuracy. The second problem is that the weakly bound states are
located far outside the range of the potential and the appropriate
long-range part of both the potential and the wave function are
therefore also decisive. A marginal numerical inaccuracy could be
crucial for the number of loosely bound states.

\subsection{Method}

Our first overall computing strategy is to use the Faddeev
decomposition of the Schr\"{o}dinger equation. The proper boundary
conditions and the more subtle correlations are then much easier to
account for in  actual computations. The second computing strategy
is to expand on a suitable angular basis for short distances and solve
the large-distance asymptotic angular equations directly at large
distances. In an overlapping region these solutions are then
combined. Finally the coupled set of radial equations is solved.

For short distances each Faddeev component is then expanded on the
hyperspherical basis \cite{lin95,zhu93,fed94a}. This amounts for each
$\rho$ to an expansion in eq.(\ref{e60}) of the functions $\phi^{(i)}$
in Jacobi polynomials. The remaining expansion coefficients are
functions of $\rho$. The total angular basis then consists of a
carefully selected set of hyperspherical harmonics with the total
angular momentum zero. The basis is characterized by the Faddeev
components $i=1,2,3$, the quantum numbers $\ell=0,1,2,3...$ and the
hyperspherical quantum number $K=0,2,4,6...$.

Only basis functions with the correct symmetry properties are
included, i.e. symmetric or antisymmetric for bosons or fermions.
When $\rho^2v_i$ is small for small $\rho$ the angular eigenvalue
equation eq.(\ref{e25}) only includes the kinetic energy operator with
the corresponding hyperspherical harmonics as solutions. In this case
a few of the lowest $\ell$ and $K$ quantum numbers are sufficient in
the basis.

For a repulsive core, where $\rho^2v_i$ diverges (or simply is very
large), larger $\ell$-values are needed in the basis to produce a
vanishing wave function at small distances. The angular matrix
elements of the potentials become large for small $\rho$ even for high
$\ell$, where the contribution usually vanishes due to the centrifugal
barrier. A faster convergence in $\ell$ is achieved by subtracting a
constant $V_0$ from each of the two-body potentials and subsequently
correcting the energy eigenvalue by $3V_0$. This is simply a shift of
the energy scale by $3V_0$.  If the infinite basis is used the choice
of constant makes no difference. A minimal basis size is obtained when
$V_0$ approximately equals the lowest angular eigenvalue.

Increasing $\rho$ normally decreases the required maximum $\ell$-value
since the centrifugal barrier then increases. On the other hand larger
hyperspherical quantum numbers (or polynomial orders) are needed,
because for short-range interactions the potentials in $\alpha$-space
become increasingly confined to a narrow region of phase space around
$\alpha_i=0$.  Accurate wave functions in this decreasing region
around $\alpha_i=0$ require high-order Jacoby polynomials (high $K$)
to reproduce the rapid change caused by the corresponding behavior of
the potential.

In the limit of very large $\rho$ the hyperspherical spectrum is also
approached for short-range potentials. Therefore in this extreme limit
only small $\ell$ and small $K$ are needed to obtain the angular
eigenvalue spectrum. However, in the sometimes very extended
transition region of intermediate $\rho$-values the eigenvalues are
required with high precision. Thus, increasing $\rho$ demands an
increasing basis for an accurate description. It is crucial that this
basis only needs to include high-order polynomials, but still only
small angular momenta. The necessary basis can then be substantially
reduced from the full hyperspherical basis with all $K$ and $\ell$ due
to the use of the Faddeev equations instead of the Schr\"{o}dinger
equation.

We shall treat large $\rho$ by considering $s$-waves, $\ell=0$,
without the further expansion on the Jacobi-polynomia, i.e. we shall
solve the large-distance equations (\ref{e93}) directly.  Moving
towards smaller $\rho$ we must then find an overlap region where the
large- and small-distance solutions coincide. To optimize
computational speed and accuracy the basis must therefore be carefully
selected to provide sufficient accuracy in the overlap region.

When the asymptotic states correspond to $\ell \neq 0$ the convergence
is expected to be much faster and essentially reached already at small
distance. We extrapolate the $\lambda$-values by using the form
$K(K+4)+c\rho^{-2}$ where $c$ is a constant obtained by matching and
$K$ is the asymptotic hyperspherical quantum number of this
eigenvalue.

The direct way of calculating the coupling terms $P_{nn'}$ and
$Q_{nn'}$ is to perform a numerical derivation, but this numerical
double derivative is inaccurate in regions where the wave functions
change rapidly, i.e. for repulsive cores at small distances. Instead
we used the explicit expressions in eq.(\ref{e49}) for small
distances.  For large distances we used numerical derivation for the
$P_{nn'}$ and $Q_{nn}$ and a smooth extrapolation of the small
distance functions for the non-diagonal $Q_{nn'}$.

\subsection{Parameters}

In the actual cases of the He-trimer systems we choose the mass unit
$m$ in eq.(\ref{e10}) in the definition of the Jacobian coordinates
\cite{fed94a,zhu93,lin95} to be one nuclear mass unit
$m=1822.887$~a.u. (atomic mass unit). The masses of the
He-atoms are then $m(^4\hbox{He})=4.002603250 m$ and
$m(^3\hbox{He})=3.016026 m$.

The two-body interaction is the LM2M2 potential from \cite{azi93}. The
three-body interaction $V_3$ is very small \cite{par94} and assumed to
be zero in the present computation.  For the $^4$He-trimer the
effective range of the potential is $R_{eff}=13.843$ a.u.  (1 atomic
length units = 0.529177 \AA) and the binding energy is estimated to be
of the order of 1mK \cite{esr96,lou96} (1 Kelvin = 0.0861735 meV =
3.16679 \E{-6} atomic energy units).  This extremely low binding
energy is equivalent to a two-body scattering length of $a_s=-189.054$
a.u., which is much larger than the effective range. The size of the
$^4$He-dimer is measured to be $\langle|{\bf r}_1-{\bf r}_2|\rangle =
120 \pm 20$ a.u. \cite{lou96}. For the $^4\hbox{He}-^3\hbox{He}$
system the different masses instead give $a_s=33.261$ a.u. and
$R_{eff}=18.564$ a.u. and for $^3\hbox{He}-^3\hbox{He}$ we get
$a_s=13.520$ a.u. and $R_{eff}=25.717$ a.u.

For the $^4$He-trimer we used a basis set of $150 ,40$ Jacobi
polynomials for each Faddeev component for $\ell=0,2$, respectively
and $30$ for $\ell= 4,6,8$.  This amounts to 840 basis states in
total, but after projecting on the states with correct symmetry the
basis size was reduced to 235.  For the $^3$He$^4$He$_2$ and
$^4$He$^3$He$_2$-molecules we used a basis set of $150 ,60 ,50, 45,
40, 35, 30, 25$ Jacobi polynomials for each Faddeev component for
$\ell=0,1,2,3,4,5,6,7$, respectively and $20$ for
$\ell=8,9,10,11,12,13,14$. The components related to two identical
$^4$He or $^3$He-atoms only needs even $\ell$ due to the symmetry
requirement. For the asymmetric trimers this amounts to a total basis
of $1500$ which reduced to 638 after symmetry restoration.  In all
cases this corresponds to an overlap region between small and
large distances around $\rho=100$~a.u. The converged results from the
basis diagonalization and the asymptotic solutions are then used below
and above 100 a.u., respectively.

The convergence and the accuracy of the numerical results were tested
thoroughly by varying the number of Jacobi polynomials in each
component as well as the number of partial waves. The results are also
independent of relatively small variations of the position of the
matching point between the small and large distance solutions. Thus we
expect reliable numerical results.

\section{Numerical results}

The numerical procedure and the input parameters are now
specified. The results are obtained by expansion on the adiabatic
basis and we shall therefore first compute the corresponding angular
eigenvalues. Then we shall show energies and wave functions for both
the LM2M2 and the simple schematic potentials. Finally we discuss the
various intrinsic geometric structures with special emphasis on the
possible appearence of Efimov states.

\subsection{Angular eigenvalues}

\begin{figure}
\begin{center}
\psfig{file=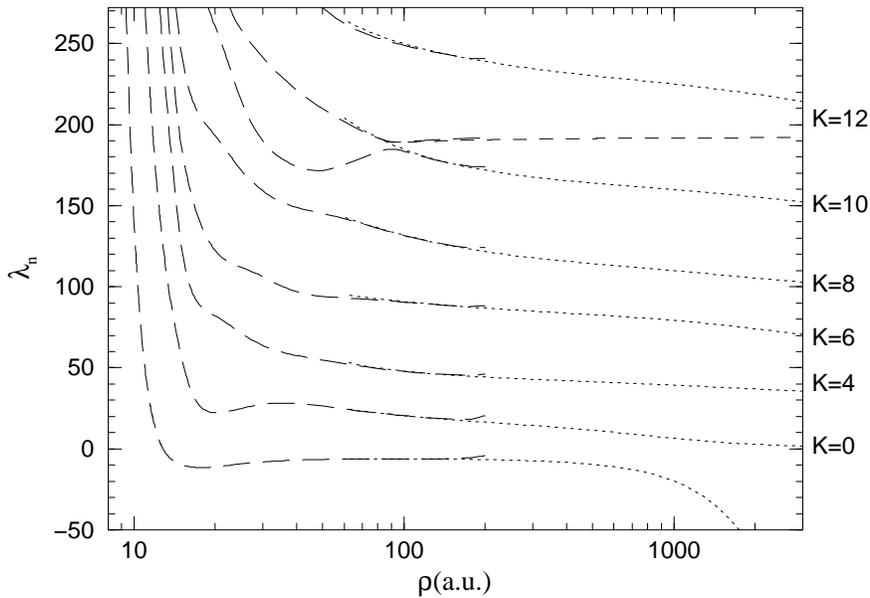,bbllx=12pt,bblly=522pt,bburx=256pt,bbury=756pt}
\end{center}
\renewcommand{\baselinestretch}{1.0}
\caption{The lowest angular eigenvalues $\lambda_n$ for
the $^4$He-trimer as functions of $\rho$. The logarithmic scale on the
$\rho$-axis uses the atomic length unit. The long-dashed curves are
the angular eigenvalues obtained by diagonalization for the relatively
small distances, the dotted curves are the large-distance $s$-wave
solution and the short dashed curve is the extrapolation of the 7'th
eigenvalue (non-zero $\ell$) obtained by using the form
$12(12+4)+c\rho^{-2}$.  The hyperspherical quantum numbers $K$ are
shown to indicate the asymptotic spectrum $K(K+4)$ for
$\rho\rightarrow\infty$.  The 7'th and 8'th eigenvalues both approach
the same asymptotic value $12(12+4)=192$. The lowest eigenvalue
diverges as $-\rho^2$ corresponding to the bound dimer state.}
\label{fig1}
\end{figure}

The decisive quantities are the adiabatic potentials where the main
components are the angular eigenvalues.  The lowest of these are
computed and shown in fig.~\ref{fig1} s functions of $\rho$ for the
$^4\hbox{He}$-trimer.  Both the general small-distance numerical
solutions and the asymptotic large-distance $s$-wave solutions are
shown.  They diverge for small $\rho$ due to the strongly repulsive
core in the interaction between two He-atoms.  The lowest of these
eigenvalues has an attractive pocket at small distance and diverges
parabolically towards $-\infty$ for large $\rho$. The pocket is
responsible for the three-body bound states and the divergence
reflects the bound state in the two-body subsystem. Due to the very
small two-body binding energy the lowest level remains almost constant
up to about 1000 a.u. where the divergence sets in.  The overlap
region above which the asymptotic solutions can be used is seen to be
around 100 a.u.

The higher lying eigenvalues asymptotically approach the
hyperspherical spectrum defined by $K(K+4)$ where $K$ is a
non-negative integer. In the present case $K$ must furthermore be an
even number because the total orbital angular momentum is zero.  The
requirement of totally symmetric wave functions removes a number of
otherwise possible degenerate states. This eliminates completely all
$K=2$ states whereas the asymptotic states corresponding to
$K=0,4,6,8,10$ all are non-degenerate and with the angular quantum
number $\ell=0$. For $K=12$ the s-wave solution is accompanied by a
higher angular momentum solution, which has converged to the
asymptotic value 12(12+4)=192 already at about $\rho=100$ a.u. This
level and that of $K=10$ avoided crossing at about $\rho=80$ a.u. The
even higher lying levels are also degenerate corresponding to more
than one totally symmetric state.

\begin{figure}
\begin{center}
\psfig{file=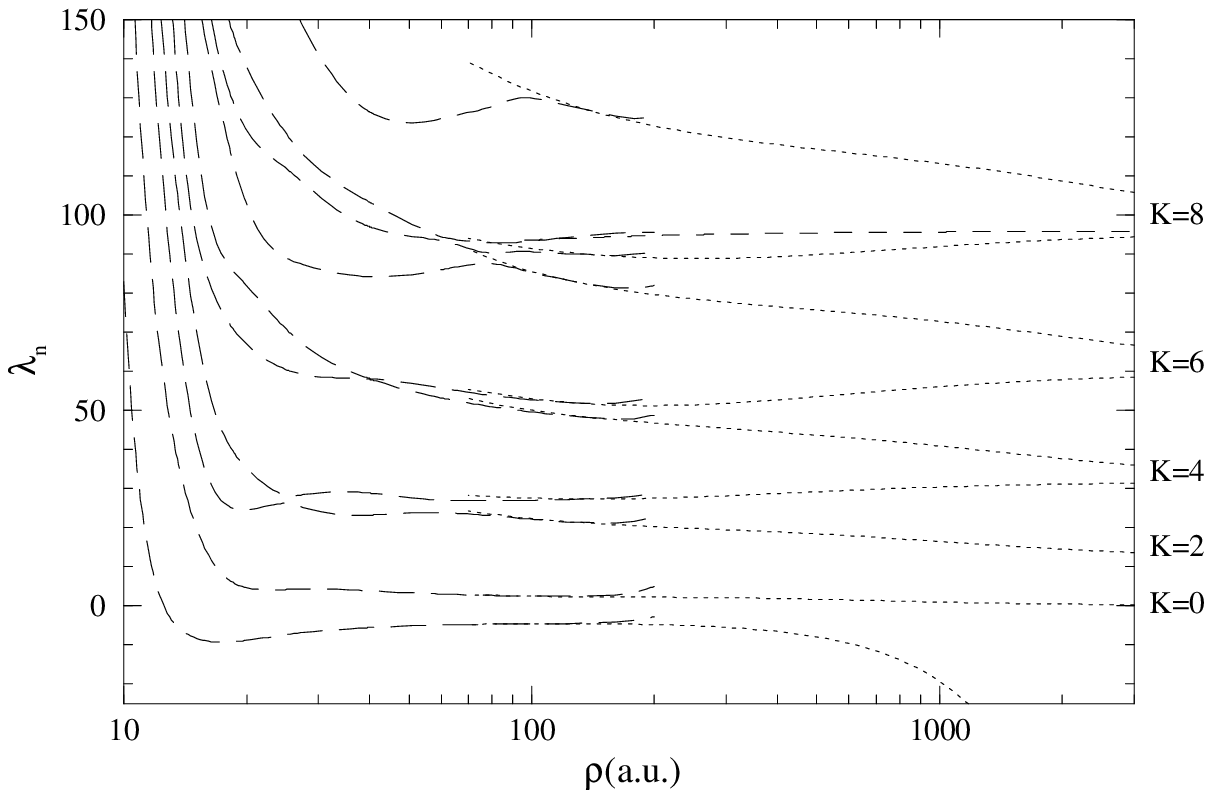,bbllx=12pt,bblly=522pt,bburx=256pt,bbury=756pt}
\end{center}
\renewcommand{\baselinestretch}{1.0}
\caption{The lowest angular eigenvalues $\lambda_n$ for
the $^3\hbox{He}^4\hbox{He}_2$-trimer as functions of $\rho$. The
logarithmic scale on the $\rho$-axis uses the atomic length unit. The
long-dashed curves are the angular eigenvalues obtained by
diagonalization for the relatively small distances, the dotted curves
are the large-distance $s$-wave solution and the short dashed curve is
the extrapolation of the 9'th eigenvalue (non-zero $\ell$) obtained by
using the form $8(8+4)+c\rho^{-2}$.  The hyperspherical quantum
numbers $K$ are shown to indicate the asymptotic spectrum $K(K+4)$ for
$\rho\rightarrow\infty$.  The lowest eigenvalue diverges as $-\rho^2$
corresponding to the bound dimer state.}
\label{fig2}
\end{figure}

The angular eigenvalue spectrum changes when the asymmetric
$^3\hbox{He}^4\hbox{He}_2$-molecule is considered, see fig.~\ref{fig2}. The
interactions are the same as for the symmetric case but the masses are
different. The number of levels is roughly increased by a factor of two
due to the less restrictive symmetry requirement. The divergence for
$\rho=0$, the parabolic divergence at large $\rho$ due to the two-body
bound state and the convergence to the hyperspherical spectrum at
large distances are still the dominating features.  Compared to the
$^4$He-system we now find a number of additional levels. They
correspond to nearly antisymmetric solutions in the exchange of $^3$He
and $^4$He whereas the ``original'' levels also present for the
$^4$He-trimer must be of symmetric character.

\begin{figure}
\begin{center}
\psfig{file=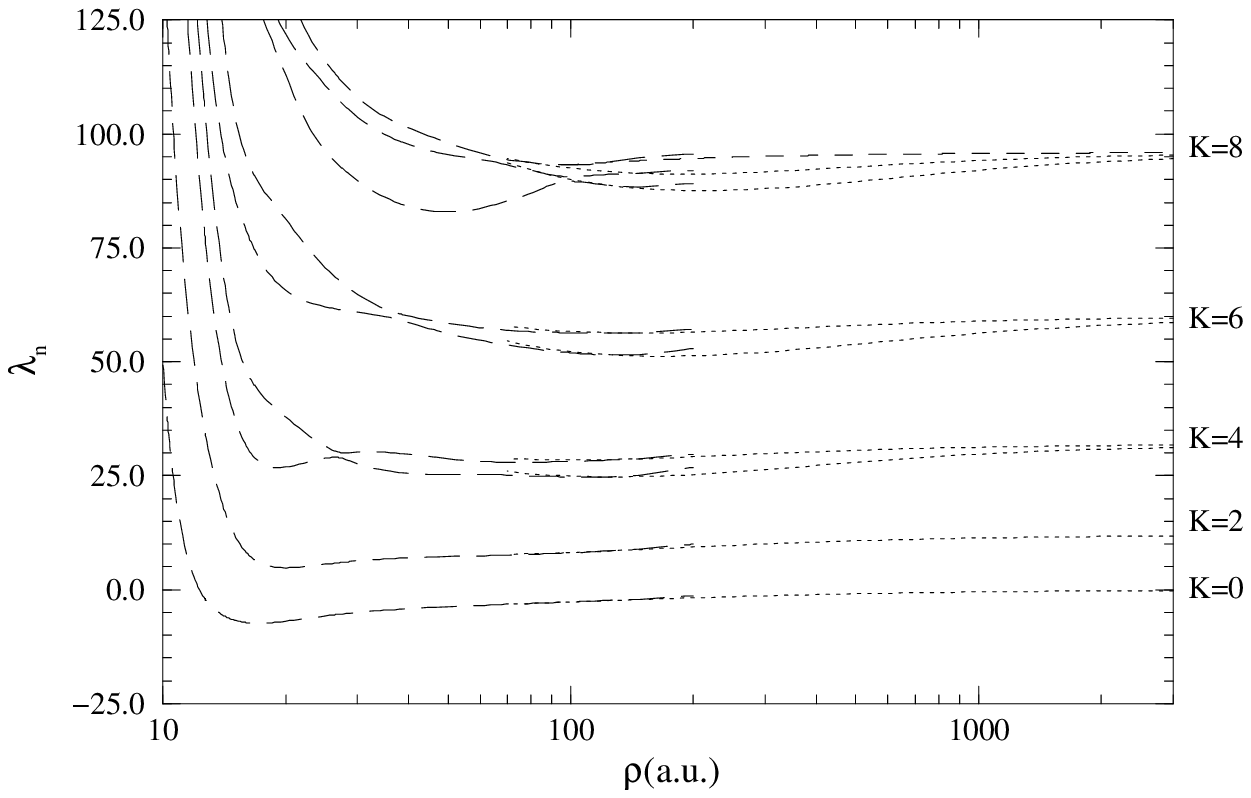,bbllx=12pt,bblly=522pt,bburx=256pt,bbury=756pt}
\end{center}
\renewcommand{\baselinestretch}{1.0}
\caption{The lowest angular eigenvalues $\lambda_n$ for
the $^4\hbox{He}^3\hbox{He}_2$-trimer as functions of $\rho$. The
logarithmic scale on the $\rho$-axis uses the atomic length unit. The
long-dashed curves are the angular eigenvalues obtained by
diagonalization for the relatively small distances, the dotted curves
are the large-distance $s$-wave solution and the short dashed curve is
the extrapolation of the 9'th eigenvalue (non-zero $\ell$) obtained by
using the form $8(8+4)+c\rho^{-2}$.  The hyperspherical quantum
numbers $K$ are shown to indicate the asymptotic spectrum $K(K+4)$ for
$\rho\rightarrow\infty$.  The lowest eigenvalue remains finite at
large $\rho$.}
\label{fig3}
\end{figure}

Considering the $^4\hbox{He}^3\hbox{He}_2$-molecule obtained by
substituting another $^4$He by a $^3$He-atom again decreases all the
binding energies. We show the angular eigenvalue spectrum in
fig.~\ref{fig3}, where the usual features are seen. However, now the
lowest eigenvalue converges to zero for large $\rho$, since none of the
two-body subsystems form bound states. The asymptotic spectrum is now
reached much faster than for the two previous cases. This rate of
convergence is roughly proportional to the sum of the three scattering
lengths or more precisely
$\frac{16}{\pi}\sum_{i=1,3}{a_s^{(i)}}\mu_{jk}$ ($a_s^{(i)}$ is the
scattering length of the two-body system with particles $j$ and $k$),
i.e.  -2042 a.u., -459 a.u. and 195 a.u., respectively. The different
rates are clearly seen in figs.~\ref{fig1}, \ref{fig2} and \ref{fig3}
especially when the range of the plotted $\lambda$-values are
considered. The levels all approach their asymptotic values from
below, since the sum of scattering lengths in this case is positive,
compare to eq.(\ref{e28}) for identical particles for $s$-waves.

\subsection{Energies and wave functions}

For halo states the decisive properties of the potentials are related
to the low-energy scattering behavior or equivalently to the large
distance behavior. This is formally expressed in the effective range
theory where disparate potentials give the same low-energy results
provided they have the same scattering length and effective
range. Since smooth and slowly varying potentials allow easy and
accurate computations we adjusted the parameters of a square well
$-S_0\Theta(r<R_0)$, a gaussian $V(r)=-S_0\exp(-r^2/b^2)$ and an
exponential potential $V(r)=-S_0\exp(-r/b)$ to reproduce $a_s$ and
$R_{eff}$ of the LM2M2-potential. The resulting parameters are given
in table \ref{pot_fit}.
\begin{table}
\renewcommand{\baselinestretch}{0.9}
\caption{Strength and range arameters for gaussian,
$V(r)=-S_0\exp(-r^2/b^2)$, exponential $V(r)=-S_0\exp(-r/b)$ and
square well $-S_0\Theta(r<b)$ potentials for the different He two-body
systems. They all have the same scattering length $a_s$ and effective
range $R_{eff}$ as the LM2M2 interaction \cite{azi93}, i.e.
$a_s=-189.054, 33.261, 13.520$ a.u. and $R_{eff}=18.564, 13.843,
25.717$ a.u. for $^4$He-$^4$He, $^4$He-$^3$He, $^3$He-$^3$He,
respectively. The strengths are in Kelvin (1 K = 0.0861735 meV =
3.16679 \E{-6} a.u.) and the ranges are in atomic units (1 a.u =
0.524177 \AA).}
\vspace{0.5cm}

\renewcommand{\baselinestretch}{1.5} \small
\begin{center}
\begin{tabular}{|c||c|c||c|c||c|c|}
\hline 
Potential & \multicolumn{2}{c||}{$^4$He-$^4$He} & 
  \multicolumn{2}{c|}{$^4$He-$^3$He} & \multicolumn{2}{c|}{$^3$He-$^3$He} \\
\hline
  & $S_0(K)$ & $b(a.u.)$ & $S_0(K)$ & $b(a.u.)$ & $S_0(K)$ & $b(a.u.)$ \\ 
\hline
 Gaussian    &  1.227    & 10.03 & 0.8925  & 10.55 & 0.16186 & 11.27 \\
 Exponential &  3.909    & 4.117 & 3.284   & 4.088 &2.499 & 4.233 \\
 Square well &  0.5578   & 14.28 & 0.3691  & 15.61 & 0.2399 & 17.07  \\    
\hline 
\end{tabular}
\end{center}
\label{pot_fit}
\end{table}

\begin{figure}
\begin{center}
\psfig{file=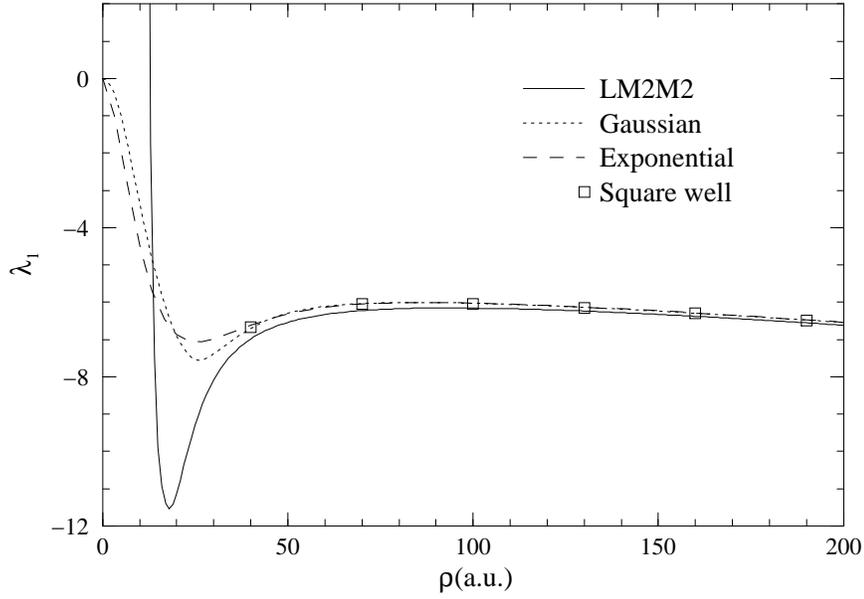,bbllx=12pt,bblly=522pt,bburx=256pt,bbury=756pt} 
\end{center}
\renewcommand{\baselinestretch}{1.0}
\caption{The lowest angular eigenvalue as a function of
$\rho$ for the $^4$He-trimer for the realistic LM2M2 \cite{azi93}
potential and the three schematic potentials defined in table
\ref{pot_fit}. The square well results are obtained from eq.(\ref{e1}).}
\label{fig4}
\end{figure}

The lowest and decisive eigenvalue for these potentials is plotted in
fig.~\ref{fig4} as function of $\rho$. The eigenvalues for all the
three simple potentials coincide for $\rho> 40$~a.u. whereas that of
LM2M2 lies slightly below and slowly approaching the others as $\rho$
increases. The deviations are enhanced due to the focus on one
$\lambda$-value and the restriction to distances less than 200 a.u.
The differences can only be interpreted as effects of terms of higher
order than in the effective range expansion. They are not completely
negligible in this region because of the large repulsive core of the
LM2M2-potential. At smaller distances the simple potentials produce
eigenvalues approaching zero for $\rho=0$. This behavior is then
compensated by less attraction in the pocket region.

The binding energies are obtained by solving the coupled set of radial
equations. The lower and upper limits can be found by using only the
lowest adiabatic potential with and without the diagonal term
$Q_{11}$, see \cite{spr67}. These limits were computed in
\cite{esr96}. We repete this computation using our method and we
furthermore increase the number of adiabatic potentials until the
results have converged. For the $^4$He-trimer we also find one excited
state in agreement with the discussion in connection with
eq.(\ref{e81}). One additional oscillation would extend beyond 10
times the scattering length and a third excited state is therefore not
possible even without the $Q_{11}$-term.

We compare in table \ref{He4results} binding energies for the LM2M2
and the simple interactions for both the ground state and the excited
state of the $^4$He-trimer.  We confirm the limits computed in
\cite{esr96}. The lowest 7 to 8 adiabatic potentials are needed to
obtain the accurate energies deviating from the upper limit by about
20\% and 7\% for the ground and excited states, respectively. This
relatively slow convergence for the LM2M2 potential is due to the
large repulsive core.  Less adiabatic potentials are needed for
convergence for the simple two-body interactions, where the lower and
upper limits in any case are very close and a larger basis therefore
also for this reason is less needed. The binding energies for the
gaussian potential are about 20\% and 10\% higher than for the LM2M2
interaction for the ground state and the excited state,
respectively. The corresponding numbers for the exponential potential
are 40\% and 20\%.
\begin{table}
\renewcommand{\baselinestretch}{0.9}
\caption{The energies in mK of the ground state and the excited state
of the $^4$He-trimer for one realistic and two schematic
potentials. The rows in the table display the numerical results for
different numbers of adiabatic channels in the basis.  We also give
the results of Esry et al. \cite{esr96} calculated for the LM2M2
potential \cite{azi93}. The row labeled $Q_{11}=0$ displays the
results obtained with the lowest adiabatic potential without the
diagonal term $Q_{11}$.}
\vspace{0.5cm}

\renewcommand{\baselinestretch}{1.1} \small
\begin{center}
\begin{tabular}{|c||c|c||c|c||c|c||c|c|}
\hline
Potential  & \multicolumn{2}{c||}{Esry et.al \cite{esr96}} & 
\multicolumn{2}{c||}{LM2M2 \cite{azi93}}   &
     \multicolumn{2}{c||}{Gaussian} &\multicolumn{2}{c|}{Exponential}\\
\hline
$Q_{11}=0$ & -293.7&-3.518 & -293.5&-3.517 & -152.6&-2.576  & -176.0&-2.823 \\
\hline
1          & -106.1&-2.118 & -105.9&-2.121 & -150.2&-2.468  & -173.9&-2.714 \\
2          &       &       & -119.5&-2.224 & -150.6&-2.482  &-174.1&-2.729 \\
3          &       &       & -122.0&-2.245 & -150.7&-2.485  &-174.2&-2.731 \\
4          &       &       & -123.9&-2.259 & -150.7&-2.485  &-174.2&-2.731 \\
5          &       &       & -124.3&-2.263 & -150.7&-2.485  &-174.2&.2.732 \\
6          &       &       & -124.7&-2.265 & -150.7&-2.486  &-174.2&-2.732 \\
7          &       &       & -125.1&-2.267 & & & &  \\
8          &       &       & -125.2&-2.269 & & && \\
\hline
\end{tabular}
\end{center}
\label{He4results}
\end{table}

The large-distance tails of gaussian and exponential two-body
potentials are very different and the variations established here
therefore measure the uncertainties inherent in such estimates.  These
results strongly indicate that such weakly bound states can be studied
by simple but carefully chosen potentials provided an accuracy of less
than about 40\% and 20\% are required for states located respectively
at the edge and far outside the short-range potential. Further
accuracy can be obtained with an appropriate choice of radial shape of
the simple two-body potential.  The computations simplify due to the
lack of repulsion at small distances and the smaller basis required.

Substituting one of the $^4$He-atoms in the $^4$He-trimer by a
$^3$He-atom decreases the binding energies and only the ground state
remains bound. We compare in table \ref{He344results} the results for
different potentials. Unlike the symmetric case we now obtain slightly
less binding energy than in \cite{esr96}. The difference is less than
0.1\% compared to the strengths of the potentials but about 5\%
compared to the binding energy itself.  Our convergence was thoroughly
tested again, but we were not able to find the origin of the
discrepancy.  As for the symmetric system an accurate binding energy
requires 5 to 6 adiabatic potentials for the LM2M2 interaction and
again about 40\% is gained compared to the upper limit obtained by
using only the lowest adiabatic potential. The simple potentials are
now off by factors 1.4 and 1.8 for gaussian and exponential radial
shapes, respectively.
\begin{table}
\renewcommand{\baselinestretch}{0.9}
\caption{The energies in mK of the ground state of the
$^3$He$^4$He$_2$-trimer for one realistic and two schematic
potentials. The rows in the table display the numerical results for
different numbers of adiabatic channels in the basis.  We also give
the results of Esry et al. \cite{esr96} calculated for the LM2M2
potential \cite{azi93}. The row labeled $Q_{11}=0$ displays the
results obtained with the lowest adiabatic potential without the
diagonal term $Q_{11}$.}
\vspace{0.5cm}

\renewcommand{\baselinestretch}{1.5} \small
\begin{center}
\begin{tabular}{|c||c||c|c|c|}
\hline
Potential & Esry et.al \cite{esr96} & LM2M2 \cite{azi93} & Gaussian 
& Exponential \\
\hline
\hline
$Q_{11}=0$& -86.52     & -86.47& -20.44   & -26.06 \\
\hline
\hline  
1         & -10.22     & -9.682& -18.41   & -24.27 \\
\hline
2         &            & -10.12& -18.67   & -24.52 \\
\hline 
3         &            & -11.90& -18.74   & -24.57 \\
\hline
4         &            & -12.74& -18.79   & -24.60 \\
\hline
5         &            & -12.86& -18.81   & -24.61 \\
\hline
6         &            & -13.23& -18.81   & -24.62 \\
\hline
7         &            & -13.34& -18.81   & -24.61 \\
\hline
8         &            & -13.55& -18.82   & -24.61 \\
\hline
9         &            & -13.64& -18.82   & -24.61 \\
\hline
10        &            & -13.66& -18.82   & -24.61 \\
\hline
\end{tabular}      
\end{center}
\label{He344results}
\end{table}

The radial wave function has a distribution of components
corresponding to the different adiabatic potentials, see
eq.(\ref{e20}). The lowest is by far carrying the largest probability
as seen in tables \ref{He4destribute} and \ref{He344destribute} for
the two He-trimer systems. The exceedingly small occupation
probability for the higher lying angular eigenvalues seems off hand to
contradict the relatively large contribution to the binding energies.
The rather instructive explanation for weakly bound systems is that
the variation in binding energy might appear to be large when compared
to the binding energy itself, but insignificant compared to the
strength of the interactions responsible for the binding, see
fig.~\ref{fig4} where the potential depths are around 2 K compared to
the binding energies of mK. Thus the lowest adiabatic potentials give
a good approximation for the wave functions although the corresponding
binding energies only are obtained with less relative accuracy.
\begin{table}
\renewcommand{\baselinestretch}{0.9}
\caption{Occupation probabilities as functions of adiabatic channels
for the ground state (left) and the excited state (right) of the
$^4$He-trimer for the three different potentials in table
\ref{pot_fit}. }
\vspace{0.5cm}

\renewcommand{\baselinestretch}{1.5} \small
\begin{center}
\begin{tabular}{|c||c|c||c|c||c|c|}
\hline
Potential & \multicolumn{2}{c||}{LM2M2 \cite{azi93}} 
& \multicolumn{2}{c||}{Gaussian}& 
           \multicolumn{2}{c|}{Exponential} \\
\hline
  1  & 0.9978      &0.9989       & 0.9997    &  0.9991   &
 0.9998  &     0.9990  \\
  2  &1.72\E{-3} &1.02\E{-3} & 2.61\E{-4}& 9.16\E{-4}&
1.69\E{-3}& 9.8\E{-4}\\
  3  &4.0\E{-4} &2.7\E{-5} & 2.8\E{-5} & 1.9\E{-5}&
1.6\E{-4} & 1.8\E{-4}\\
  4  &8.5\E{-5} &5.5\E{-6} & 1.5\E{-6} & 4.3\E{-6}&
8.9\E{-7} &3.7\E{-5}\\
  5  &2.4\E{-5} &2.6\E{-6} & 5.9\E{-7} & 2.0\E{-6}&
3.7\E{-6} &8.4\E{-7}\\
  6  &4.6\E{-6} &8.3\E{-7} & 1.4\E{-7} & 5.8\E{-7}&
9.4\E{-8}&2.6\E{-7}\\
  7  &1.0\E{-5} &1.2\E{-7} & &&&\\
  8  &3.9\E{-6} &3.2\E{-7} & &&&\\
\hline 
\end{tabular}
\end{center}
\label{He4destribute}
\end{table}
\begin{table}
\renewcommand{\baselinestretch}{0.9}
\caption{Occupation probabilities as functions of adiabatic channels
for the ground state of the $^3$He$^4$He$_2$-trimer for the three
different potentials in table \ref{He344results}. }
\vspace{0.5cm}

\renewcommand{\baselinestretch}{1.5} \small
\begin{center}
\begin{tabular}{|c||c|c|c|}
\hline
Potential & LM2M2 & Gaussian   & Exponential \\
\hline
  1  & 0.9959    & 0.9974    & 0.9979 \\
  2  &3.43\E{-3}&2.50\E{-3}&2.03\E{-3}\\
  3  & 2.8\E{-4}& 1.0\E{-4}& 7.5\E{-5}\\
  4  & 2.4\E{-4}& 2.6\E{-5}& 1.2\E{-5}\\
  5  & 6.9\E{-5}& 1.6\E{-5}& 9.9\E{-6}\\
  6  & 4.7\E{-5}& 5.4\E{-7}& 3.0\E{-7}\\
  7  & 1.9\E{-6}& 8.9\E{-7}& 5.7\E{-7}\\
  8  & 1.9\E{-5}& 1.2\E{-6}& 6.9\E{-7}\\
  9  & 2.4\E{-6}&           &\\
 10  & 6.1\E{-7}&           &\\
\hline
\end{tabular}
\end{center}
\label{He344destribute}
\end{table}

\begin{figure}
\begin{center}
\psfig{file=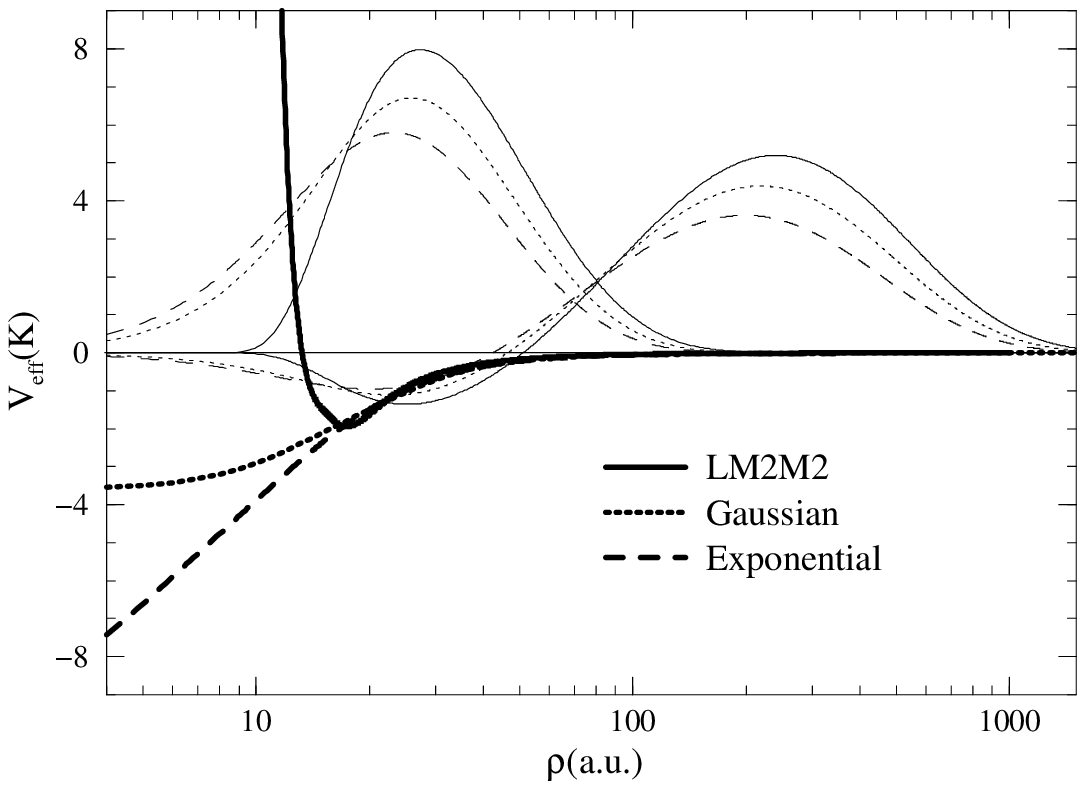,bbllx=12pt,bblly=522pt,bburx=256pt,bbury=756pt} 
\end{center}
\renewcommand{\baselinestretch}{1.0}
\caption{The lowest adiabatic effective radial potentials
(thick curves) $V_{eff}(\rho)={\hbar^2\over 2m}
\left((\lambda_1+\frac{15}{4})\rho^{-2} - Q_{11}\right)$ and the wave
functions corresponding to the two bound states (thin curves) for the
$^4\hbox{He}$-trimer for one relalistic \cite{azi93} and two schematic
two-body interactions defined in table \ref{pot_fit}.  The scale on
the $\rho$-axis is logarithmic.}
\label{fig5}
\end{figure}

The dominating components of the radial wave functions for the ground
state and excited state of the $^4$He-trimer are shown in
fig.~\ref{fig5} along with the different effective radial
potentials. The repulsive core pushes the ground state towards larger
distances, whereas the simple potentials allow finite probability also
at small distances. The excited state is already at a large distance
and therefore less affected by the changing potentials.  The wave
functions related to the different potentials are quite similar.  The
length of the first oscillation is about the 350 a.u. as predicted
from eq.(\ref{e79}) with $R_{eff}=18$ a.u. The next would extend 20
times as far, i.e. about 7000 a.u.

\subsection{Geometric structure}

The wave functions in fig.~\ref{fig5} are clearly not revealing
directly the geometric structure of these three-body systems. This is
not only because the angular part of the wave function is missing, but
also because the plotted total wave function fully includes all the
appropriate symmetries, i.e. symmetric in exchanges of identical
bosons and rotational symmetry averaging over all directions. To get a
visual picture of the structure we want to consider the plane through
the three particles at points ${\bf r}_i$. The center of mass is at
the origin. Let us take the y-axis as the principle axis with the
lowest moment of inertia. We furthermore assume that one particle is
in the first and two particles are in the third and fourth
quadrants. This leaves us with the intrinsic geometry of the trimer.

\begin{figure}
\begin{center}
\psfig{file=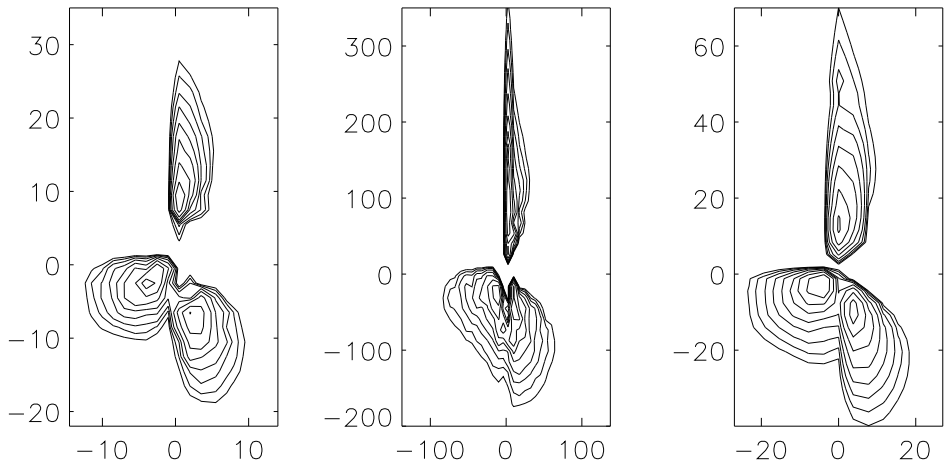,bbllx=80pt,bblly=370pt,bburx=500pt,bbury=550pt} 
\end{center}
\renewcommand{\baselinestretch}{1.0}
\caption{The contour diagrams of the particle density distributions
in the intrinsic coordinate system for the ground state (left) and the
excitede state of the $^4\hbox{He}$-trimer (middle) and the bound
state of $^3\hbox{He}^4\hbox{He}_2$ (right).  The units on the axes
are atomic lengths units. They are different for the three states.
There is a factor of 1.5 between each contour curve. The scales on the
plots are different.}
\label{fig6}
\end{figure}

The resulting contour plots are shown in fig.~\ref{fig6}. Note that
the distance scales on the individual figures are quite different. The
triangular shape for the $^4$He-trimer ground state (left) is quite
apparent. The excited state in the middle show that two of the
particles prefer to be close while the third particle is further away.
The Efimov conditions are almost fulfilled, i.e. very large scattering
lengths for at least two of the binary subsystems.  As we shall
discuss later this figure illustrates the structure of an Efimov
state.  Any higher-lying Efimov state could simply be found by a
scaling of the axes, see eq.(\ref{e78}).

The contour diagram at the right side of fig.~\ref{fig6} also show the
preference for two relatively close-lying particles. However, only one
scattering length is now large and the Efimov conditions are not
fulfilled. It is instructive to imagine that the $^3$He-$^4$He
scattering length is increased and the subsequent appearence of the
Efimov states. The lowest eigenvalue in fig.~\ref{fig2} would still
bend over and diverge due to the bound state. The second eigenvalue
would approach a negative constant above the lowest eigenvalue. The
corresponding adiabatic potential would isolated provide the Efimov
states. However, these states would appear in the continuum as excited
states above the ground state shown in the figure. Thus the ground
state of the asymmetric system cannot be an approximate Efimov state.
The superficial resemblance of the middle and right figures therefore
illustrates how deceiving the density distribution can be. Additional
information is needed to come up with the correct interpretation.

\begin{figure}
\begin{center}
\psfig{file=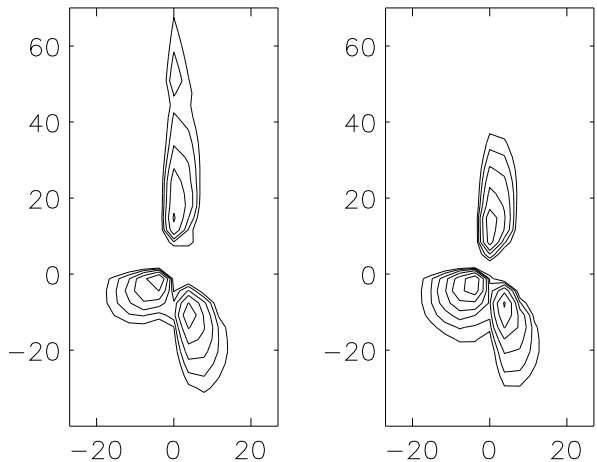,bbllx=80pt,bblly=370pt,bburx=500pt,bbury=550pt}
\end{center}
\renewcommand{\baselinestretch}{1.0}
\caption{The contour diagram of the $^3$He density (left) and half of
the $^4$He density distribution (right) in the intrinsic coordinate
system for the bound state of $^3\hbox{He}^4\hbox{He}_2$ (right). The
units on the axes are atomic lengths. There is a factor of 1.5 between
each contour. Both the scale and the contours are the same as on
the total density distribution in figure \ref{fig6}.}
\label{fig7}.
\end{figure}

For the asymmetric system we can also compute the individual particle
density as shown in fig.~\ref{fig7}. The left figure can be obtained
from the right figure if a small component with an elongated
distribution is added. The $^4$He-trimer appears in a completely
symmetric state while the asymmetric system in addition also has a
relatively small asymmetric component of the $^3$He-$^4$He relative
wave function. The large symmetric component resembles the
$^4$He-trimer and the small asymmetric configuration, which becomes
dominating at larger distances where the structure of the $^4$He-dimer
is approached.

The triangular structure at the left side of fig.~\ref{fig6} is apart
from the scale similar to the three-body cluster approximation of the
ground state of $^{12}$C, whereas the two other trimer states are very
different from the almost linear configuration of the astrophysically
interesting first excited $0^+$-state of $^{12}$C, which therefore is
far from being an Efimov state, see \cite{fed96}.

The geometry of the He-trimers can be studied quantitatively by
computation of various expectation values related to the distances
between the particles. We first consider the sizes shown in table
\ref{sizes} where two measures are used, i.e. the root mean square
radius and the (first order) distance from the center of mass. The
dimer has an average distance between the particles of about 100 a.u.,
which is at the lower end of the experimentally allowed range of $120
\pm 20$ a.u. \cite{lou96}. The three-body ground state is much smaller
than the $^4$He-dimer \cite{esr96,lou96}. However, the size of about
100 a.u. of the excited state is around 50\% larger than for the dimer
both when measured by the first and the second moment.  This radius is
larger than the range of the effective two-body interaction but
smaller than the size (189 a.u.) of the scattering length. The spatial
extension of the probability reaches beyond $|a_s|$ as seen in
fig.~\ref{fig5}, but the wave function is still well within the range
of the effective radial three-body potential ($\rho^{-2}$ behavior)
extending to about $|15\mu_{ij}a_s|$, see eq.(\ref{e28}). These
proportions are characteristic for an Efimov state.
\begin{table}
\renewcommand{\baselinestretch}{0.9}
\caption{Expectation values of various operators related to the
geometrical structure of the bound states of the He-dimer and the
He-trimers. The results for the different potentials in tables
\ref{He4results} and \ref{He344destribute} are given for each
quantity: LM2M2 (top), gaussian (middle) and exponential (bottom).
The coordinate of particle $i$ is ${\bf r}_i$, the center of mass
coordinate is ${\bf R}$ and $n=2,3$ is the number of particles.  All
the lengths are in atomic units.}
\vspace{0.5cm}
\renewcommand{\baselinestretch}{1.5} \small
\begin{center}
\begin{tabular}{|c|c|c|c|c|c|}
\hline
  & $^4\hbox{He}_2$  & $^4\hbox{He}_3$ & $^4\hbox{He}_3^*$ & 
$^3\hbox{He}^4\hbox{He}_2$ \\
\hline
  &  66.9      &     11.8    & 115       & 26 \\
$\sqrt{\left<\sum_i({\bf r}_i-{\bf R})^2\right>/n}$ 
  &  66.9      &     10.8    & 104       & 22 \\
  &  66.9      &      9.9    &  95       & 20 \\
\hline
  &  49.0      &     10.2    &  96       & 20 \\
$\left<\sum_i \left|{\bf r}_i-{\bf R}\right|\right>/n$ 
  &  49.0      &      9.4    &  86       & 17.5 \\
  &  49.0      &      8.6    &  77       & 15.6 \\
\hline
  &  98.1      &     11.8    &  70       & 20 \\
$\left<\min_{i\not=j}|{\bf r}_{ij}|\right>$ 
  &  98.0      &     10.6    &  67       & 19.1 \\
  &  98.0      &      9.6    &  64       & 17.0 \\
\hline
  &  98.1      &     23      &  219      & 46 \\
$\left<\max_{i\not=j}|{\bf r}_{ij}|\right>$ 
  &  98.0      &     21      &  195      & 40 \\
  &  98.0      &     19.3    &  178      & 35 \\

\hline
  &    1       &     0.51    &   0.34    & 0.46 \\
$\left<{\min_{i\not=j}|{\bf r}_{ij}|\over\max_{i\not=j}|{\bf r}_{ij}|}\right>$ 
  &    1       &     0.51    &   0.37    & 0.49 \\
  &    1       &     0.51    &   0.38    & 0.49 \\
\hline
\end{tabular}
\end{center}
\label{sizes}  
\end{table}

The ratio between the three-body radii of excited and ground state is
9.5 which only is two times smaller than the asymptotic prediction
from eq.(\ref{e79}).  In comparison the ratio of the binding energies
is 52 which is about 7 times smaller than predicted in eq.(\ref{e83})
for two Efimov states. These observations simply confirm that the
ground state does not have the characteristics of an Efimov state. An
additional (second) excited state (which is impossible for this
scattering length) would have obeyed these scaling laws for radii and
energies for Efimov states. 

The asymmetric system $^3$He$^4$He$_2$ has only one bound state with
both binding energy and size in between the corresponding values for
the ground state and the excited state of the symmetric system, see
tables \ref{He4results}, \ref{He344results} and \ref{sizes}.  For
Efimov states the product of the energy and the square of the radius
should be a constant, see eqs.(\ref{e79}) and (\ref{e83}). This
product is four times larger for the excited symmetric state than for
the asymmetric state.  The radius should then be about twice as large
if the state should resemble an Efimov state.  However, this seems to
be unlikely, since the radius is of the same order as the two-body
effective ranges. In any case the interpretation as an approximate
Efimov state was already excluded in the discussion above. Thus a
pronounced halo structure is therefore a more appropriate description.

The distance between the two $^4$He-particles computed as
$\langle|{\bf r}_1- {\bf r}_2|\rangle$ is 28 a.u. and the distance
between $^3$He and one specific of the $^4$He-particles is 38 a.u. The
distances between both the $^3$He-$^4$He and the two $^4$He particles
are then two to three times smaller in this asymmetric bound state
than in the bound $^4$He dimer. As necessary both distances are in
between the minimum and maximum of 20 and 46 a.u. Again this
demonstrates that the bound dimer-state is not a dominating
configuration in the three-body systems.

Another interesting quantity in table \ref{sizes} is the expectation
value of the ratio between the smallest and the largest interparticle
distance. First the minimum and maximum average distances themselves
differ by a factor of two for the ground states and a factor of three
for the excited Efimov-like state.  The ratio is nearly 0.5 for both
ground states and 0.34 for the excited state. If the angular part
(including the $\alpha$ coordinate) of the wave function was constant
this number would be $0.52$. In the case of an Efimov state this
number was computed to be $0.38$. Thus the excited state does also in
this respect resemble an Efimov state. That the ratio is less than the
value for an Efimov state indicates non-negligible contributions
corresponding to two particles in a bound state and the third particle
further away. This is consistent with the observation that the excited
stated is larger than the dimer.

In all cases the simple potentials produce sizes deviating about 10\%
and 20\% for gaussian and exponential potentials from those of the
LM2M2 interaction. Thus if only this accuracy is required the
corresponding simpler computations are sufficient.

\section{Summary and conclusions}

Spatially extended three-body halo states are apparently present in
some of the atomic helium-trimer systems. The $^4$He-dimer is
experimentally established as exceedingly large and as a consequence
it must also be very loosely bound. The large distances are then
essential in any accurate description. The possibility of finding
Efimov states is adding interest in obtaining a quantitative
understanding, but unfortunately at the same time emphasizing the
practical difficulties. The necessary large-distance correlations must
be treated very accurately without loosing the precision at smaller
distances where the binding energy originates. This three-body problem
is of quantum mechanical nature since the probability almost entirely
is found in the classically forbidden region.

We incorporate the correlations by solving the coordinate space
Faddeev equations. We expand the wave function in a complete set of
angular eigenfunctions obtained by solving the angular part of the
Faddeev equations for a given hyperradius. This adiabatic basis is
related to a set of effective radial potentials. The corresponding
radial wave functions are solutions to the coupled set of radial
differential equations. We first formulate this prescription in
general and then the crucial large distances are specifically
considered. Only $s$-waves couple at large distances and this fact is
exploited to derive simple equations with almost analytic solutions.
A suitable numerical procedure follows from this formulation. For
exploratory calculations we furthermore suggest a simplifying
approximation where the ``correct'' potential is substituted by almost
schematic potentials with relatively little loss of accuracy for the
spatially extended states.

The lowest effective radial potential is decisive and we describe a
series of its large-distance analytic properties for a symmetric
system. When the two-body scattering length is sufficiently large the
Efimov states appear as solutions. We repete the occurrence conditions
and a few of the analytic properties of these exotic states.

The strongly repulsive core is necessary in a correct description of
the two-body potential. The numerical procedure is somewhat delicate
and a sizable basis is necessary. Each of the Faddeev components is
expanded on the hyperspherical basis. We need relative angular
momentum quantum numbers up to about $10$, Jacobi polynomials for
$s$-waves up to order 150 and significantly fewer for the higher
partial waves.  This numerical procedure produces the solutions at
small distances up to about 100 a.u. and from then on we use the
solutions obtained directly from the decoupled large-distance set of
equations.

The schematic two-body potentials are parametrized to reproduce the
scattering length and effective range of the original complicated
potential. We use radial shapes of gaussian, exponential and square
well form. The solutions are much easier to compute due to the lack of
hard core repulsion. Still these simple attractive potentials
reproduce the results of the ``correct'' calculations within 40\% and
significantly better for the spatially extended states. More suited
shapes could certainly reduce this uncertainty, which in fact is
remarkable for several reasons. First this demonstrates the importance
of the large-distance behavior of the effective radial
potential. Second the error in binding energy by using such simple
potentials is roughly the same as the results obtained by using only
the lowest adiabatic potential, which carries more than 99\% of the
probability in a fully converged calculation. Thus simple potentials
could be exploited to gain even semi quantitative insight.

We have essentially confirmed the calculations of \cite{esr96} using
the lowest adiabatic potential. In our full calculations we find
additional binding of 7\% to 40\% measured relative to the binding
energy, which is several orders of magnitude smaller than the strength
of the attraction. Indeed these states are very weakly bound. We
investigated the geometry of the bound He-trimer states.  The ground
state of $^4$He$_3$ is an equal sided triangle. The excited state of
$^4$He$_3$ and the ground state of the asymmetric $^3$He$^4$He$_2$
system are both also triangular but with sides differing by a factor of
two to three.

The excited state of $^4$He$_3$ has all the characteristics of an
Efimov state whereas the somewhat similar density distribution of the
asymmetric state rather should be characterized as a pronounced halo
state.  More Efimov states would only appear in the symmetric system
when the scattering length is increased by about one order of
magnitude. Smaller changes would be needed to promote the halo state
to the giant halo state called an Efimov state.

In conclusion, we have described a method to compute spatially
extended three-body structures. The essential large-distance behavior
is specially treated. We apply the method to the atomic helium
trimers.  The binding energies are accurately determined and the
geometric structures are found to be triangular configurations.  The
ground state of $^4$He$_3$ has a triangular shape and the excited
state is an Efimov state whereas the $^3$He$^4$He$_2$ bound state is a
halo state.

\end{document}